# Thermodynamic modeling of the Pd-Zn system with uncertainty quantification and its implication to tailor catalysts

Rushi Gong[1], Shun-Li Shang[1], Hui Sun[1], Michael J. Janik[2], and Zi-Kui Liu[1]

[1]Department of Materials Science and Engineering, The Pennsylvania State University, University Park, PA 16802, USA

[2]Department of Chemical Engineering, The Pennsylvania State University, University Park, PA 16802, USA

**Abstract**

Pd-Zn intermetallic catalysts show encouraging combinations of activity and selectivity on well-defined active site ensembles. Thermodynamic description of the Pd-Zn system, delineating phase boundaries and enumerating site occupancies within intermediate alloy phases, is essential to determining the ensembles of Pd-Zn atoms as a function of composition and temperature. Combining the present extensive first-principles calculations based on density functional theory (DFT) and available experimental data, the Pd-Zn system was remodeled using the CALculation of PHAse Diagrams (CALPHAD) approach. High throughput modeling tools with uncertainty quantification, i.e., ESPEI and PyCalphad, were incorporated in the phase analysis. The site occupancies across the γ phase composition region were given special attention. A four-sublattice model was used for the γ phase owing to its four Wyckoff positions, i.e., the outer tetrahedral (OT) site 8c, the inner tetrahedral (IT) site 8c, the octahedral (OH) 12e, and the cuboctahedral (CO) site 24g. The site fractions of Pd and Zn calculated from the present thermodynamic model show the occupancy preference of Pd in the OT and OH sublattices in agreement with experimental observations. The force constants obtained from DFT-based phonon calculations further supports the tendency of Pd occupying the OH sublattice compared with the IT and CO sublattice. The catalytic assembles changing from Pd monomers ($Pd_1$) to trimers ($Pd_3$) on the surface of γ phase are attributed to the increase of Pd occupancy in the OH sublattice.

**Highlights**

- Thermodynamic re-modeling of the Pd-Zn system with uncertainty quantification (UQ)
- A 4-sublattice model used to model γ phase and its site occupancy
- Phonon calculations to understand site occupancy in γ phase
- Surface construction in γ phase examined to tailor its catalytic performance

## 1 Introduction

Palladium (Pd) alloys are known for their excellent catalytic properties in a variety of reactions such as hydrogenation of alkynes (including acetylene) and butadiene.[1–3] By alloying with inert components such as zinc (Zn), catalytic ensembles on the surface of Pd can be controlled to enhance selectivity for semi/partial-hydrogenation reactions[2,4] The Pd-Zn catalyst is also effective for other reactions, for example, methanol steam reforming and ester hydrogenation.[5,6]

In the Pd-Zn system, site occupancy of Pd and Zn in the gamma ($\gamma$) phase offers distinct advantages for precisely controlling the composition of active sites.[7] In collaboration with experimental efforts in the Rioux group,[4] we have recently illustrated that subtle changes in composition within the Pd-Zn $\gamma$ phase can be used to control the surface active site between isolated $Pd_1$ and $Pd_3$ species isolated by surrounding Zn atoms. The $\gamma$ phase has the $\gamma$-$Cu_5Zn_8$ structure with space group $I\bar{4}3m$ and 52 atoms per crystallographic cell in four Wyckoff sites, i.e., the outer tetrahedral (OT) site 8c, the inner tetrahedral (IT) site 8c, the octahedral (OH) 12e, and the cuboctahedral (CO) site 24g.[8] Adjusting compositions in the $\gamma$ phase leads to different surface chemistry for given Miller indices, as variation in elemental site occupancies exposes different Pd-Zn arrangements on the surface. Before surface site structure can be considered, a thermodynamic description of the bulk intermetallic phase is needed. Thermodynamic description of the $\gamma$ phase in Pd-Zn is hence fundamental to evaluate phase stability, site occupancy, and in turn, surface constructions, which would be helpful to design active ensembles and improve selectivity for catalytic reactions.

Thermodynamic modeling of the Pd-Zn system was performed in 2006 by Vizdal et al.[9] However, a two-sublattice model of $(Pd,Zn)_2(Pd,Zn)_9$ was used to depict the $\gamma$ phase,[9] which is inconsistent with the aforementioned four Wyckoff positions and hence hinders the analysis of surface constructions as a function of composition. In addition, enthalpy of formation ($\Delta_f H$) of the γ phase measured by Amore et al.[10] was not considered in the previous modeling.[9] The $\Delta_f H$ values of the endmember compounds in γ phase modelled by Vizdal et al.[9] were thus less reliable due to the lack of experimental data or theoretical predictions.

In the present work, thermodynamic description of the Pd-Zn system was re-modelled using the CALPHAD (CALculation of PHAse Diagram) approach[11–13] with new experimental data in the literature and first-principles calculations based on density functional theory (DFT); an improved description of thermodynamic properties is achieved. In particular, a four-sublattice model is implemented for the γ phase based on its four Wyckoff sites with the Gibbs energy of formation for each endmember predicted by DFT-based phonon calculations. The elementary interactions that dictate site occupancies is analyzed using force constants from phonon calculations. The open source software ESPEI (Extensible Self-optimizing Phase Equilibria Infrastructure)[14] with the computational engine of PyCalphad[15] is used for the evaluation of model parameters and uncertainty quantification (UQ).[16] Finally, the relationship between site occupancy and catalytic ensembles on the surface of the γ phase is rationalized.

## 2  Review of thermodynamic properties in the Pd-Zn system

The Pd-Zn system contains 3 solution phases, i.e., Liquid, FCC, and HCP, and 6 intermetallic compounds, i.e., BCC_B2 ($\beta$), FCC_L1$_0$ ($\beta_1$), Gamma ($\gamma$), Pd$_2$Zn, PdZn$_2$, and Pd$_9$Zn$_{91}$ based on the works summarized by Vizdal et al..[9] Details of these phases can be seen in Table 1, including phase name, crystallographic information, and the sublattice models of phases used in the present work.

Phase equilibrium properties of the Pd-Zn system published before 2006 were reviewed by Vizdal et al.[9] based on the previous assessments[17,18]. Massalski[17] reported the solubility of Zn in FCC-Pd to be about $x_{Zn}$ = 0.18-0.20, and the similar values were also observed by Hansen and Anderko.[18] The maximum solubility of Zn in FCC Pd was around $x_{Zn}$ = 0.26 at 1000°C reported by Chiang et al.[19] and Kou and Chang.[20] The solubility of Pd in HCP-Zn was reported to be lower than $x_{Zn}$ = 0.01.[9,17,18] Vizdal et al.[9] measured the temperatures of the invariant reactions, i.e., Liquid → Pd$_9$Zn$_{91}$ + HCP at 690 K and Liquid + Gamma → Pd$_9$Zn$_{91}$ at 707 K, in the Zn-rich region using differential thermal analysis (DTA).

Thermochemical measurements for the Pd-Zn system are scarce. Kou and Chang[20] measured the activities of Zn ($a_{Zn}$) in $\beta_1$, showing that $\ln a_{Zn}$ increases from -9.146 to -1.464 with $x_{Zn}$ from 0.3762 to 0.5779. They also reported the enthalpies of formation of $\beta_1$ as -66.5 kJ/mol-atom at 1000°C. Chiang et al.[19] measured the vapor pressures of Zn between 750 and 1300 K with $x_{Zn}$ = 0 - 0.83, using the isopiestic method. From these measurements, they determined the activities of Pd and Zn in FCC, $\beta$, and $\beta_1$ and partial molar Gibbs energy and enthalpy in $\beta_1$.[19] According to Chiang et al.,[19] at 1273 K, the activity values of $\ln a_{Zn}$ increase from -15.24 to -1.48 with increasing $x_{Zn}$ from 0.01 to 0.6; and the partial molar Gibbs energy and enthalpy reach the lowest values at $x_{Zn}$=0.5, which are -50.8±2.0 kJ/mol-atom and -70.2±10.0 kJ/mol-atom, respectively. Amore et al.[10] used calorimetry to obtain the enthalpy of formation between -33.7 and -35.1 kJ/mol-atom for the alloys with $x_{Zn}$ = 0.77 - 0.8. Site occupancy in γ were reported by Edström et al.,[8] Gourdon et al.,[21] and Dasgupta et al.[4] using X-ray diffraction (XRD). The OT sites were fully occupied by Pd, the OH sites were occupied by Pd and Zn, the IT and CO sites were occupied by Zn.

## 3 Methodology

### 3.1 DFT-based first-principles calculations

*3.1.1 Free energy at finite temperatures*

The Helmholtz energy of a given structure, $F(V,T)$, is obtained as a function of volume ($V$) and temperature ($T$) in terms of the quasiharmonic approach by DFT-based first-principles calculations and expressed as[22]

$$F(V,T) = E(V) + F_{el}(V,T) + F_{vib}(V,T) \qquad Eq.\ 1$$

where $E(V)$ is the static energy at 0 K without the zero-point vibrational energy. The equilibrium results at zero external pressure ($P = 0$ GPa) including static energy $E_0$, volume ($V_0$), bulk modulus ($B_0$) and its pressure derivate ($B'$) were obtained using a four-parameter Birch-Murnaghan (BM4) equation of state (EOS) as shown below[22]

$$E(V) = a + bV^{-2/3} + cV^{-4/3} + dV^{-2} \qquad Eq.\ 2$$

where *a, b, c,* and *d* are fitting parameters. In the present work, DFT calculations were performed with 8 volumes for EOS fitting with $V/V_0$ in the range of 0.91 ~ 1.12. The second term in ***Eq. 1***, $F_{el}(V,T)$, represents the temperature-dependent thermal electronic contribution described by,[23]

$$F_{el}(V,T) = E_{el}(V,T) - T \cdot S_{el}(V,T) \qquad Eq.\ 3$$

where $E_{el}$ and $S_{el}$ are the internal energy and entropy of thermal electron excitations, respectively, obtained from the electronic density of states (DOS). The third term in ***Eq. 1***, $F_{vib}(V,T)$, represent the vibrational contribution given by,[23,24]

$$F_{vib}(V,T) = k_B T \sum_q \sum_j \ln\left\{2\sinh\left[\frac{\hbar\omega_j(q,V)}{2k_B T}\right]\right\} \qquad Eq.\ 4$$

where $\omega_j(q,V)$ represents the frequency of the $j^{th}$ phonon mode at wave vector $q$ and volume $V$, and $\hbar$ the reduced Plank constant. The phonon DOS's for the phases of interest were calculated for at least 4 volumes, which cover the temperature range from 0 K to 1000 K.

### 3.1.2 Details of first-principles calculations

DFT-based first-principles and phonon calculations were performed to obtain Helmholtz energies of intermetallic compounds and endmembers at finite temperatures, which equal to Gibbs energies under ambient pressure. The Vienna *Ab initio* Simulation Package (VASP)[25] was employed for all DFT-based calculations. The projector augmented-wave method (PAW) was used to account for electron-ion interactions in order to increase computational efficiency compared with the full potential methods.[26,27] Electron exchange and correlation effects were described using the generalized gradient approximation (GGA) as implemented by Perdew, Burke, and Ernzerhof (PBE).[28] The GGA includes the electronic density and its gradient as exchange-correlation functionals. Furthermore, the hybrid exchange-correlation functional HSE06[29] was applied to calculate the enthalpy of formation with higher accuracy. The plane-wave basis cutoff energy was 277 eV for relaxations and 520 eV for the final static calculations. The convergence criterion of the electronic self-consistency was set as $10^{-6}$ eV/atom for relaxations and static calculations.

Table 2 provides parameters for first-principles and phonon calculations, including reciprocal k-points meshes and supercell sizes for compounds $Pd_2Zn$, $PdZn$, $PdZn_2$, 16 endmembers of γ phase (26 atoms for each endmember in the primitive cell of γ phase), where endmember represents configuration where only one component in each sublattice,[30] and $Pd_8Zn_{44}$ (the key endmember of γ phase in its crystallographic cell). The phonon calculations were performed using the supercell method. The phonon DOS's and force constants were analyzed using the YPHON code.[31] Note that Table 2 includes the supercell sizes and k-points meshes for phonon calculations, while the plane-wave cutoff energy of 277 eV was used for phonon calculations.

## 3.2 Thermodynamic modeling

In the present work, nine phases, i.e., Liquid, FCC, HCP, $β$, $β_1$, γ, $Pd_2Zn$, $PdZn_2$ and $Pd_9Zn_{91}$, are modeled based on available experimental data reviewed in Sec.2 and thermochemical data from DFT-based first-principles and phonon calculations showed in Sec.3.1. The Gibbs energy functions of pure Pd and Zn are taken from the Scientific Group Thermodata Europe (SGTE) pure element database.[32]

Gibbs energies of the solution phases $θ$ of Liquid (L), FCC, and HCP are formulated as,

$$G_m^θ = \sum_{i=Pd,Zn} x_i {}^oG_i^θ + RT \sum_{i=Pd,Zn} x_i \ln x_i + {}^{xs}G_m \qquad Eq.\ 5$$

where $x_i$ is the mole fraction of component $i$, ${}^oG_i^θ$ the Gibbs energy of component $i$, $R$ the gas constant, $T$ the temperature, and ${}^{xs}G_m$ the excess Gibbs energy. The first term represents mechanical mixing of the endmembers, here the pure elements. The second term represents the ideal configurational entropy of mixing. The third term represents the excess Gibbs energy, which is described by the Redlich-Kister polynomial,[33]

$$^{xs}G_m = x_{Pd}x_{Zn} \sum_{k=0} {}^kL_{Pd,Zn}(x_{Pd} - x_{Zn})^k \qquad Eq.\ 6$$

where $^kL_{Pd,Zn}$ is the $k^{th}$ interaction parameter between Pd and Zn, modeled as

$$^kL_{Pd,Zn} = A + BT + CT \ln T \qquad \text{Eq. 7}$$

where $A$, $B$, and $C$ are model parameters to be evaluated.

The BCC_B2 $\beta$ phase (space group $Pm\bar{3}m$ with two Wyckoff sties 1a and 1c) appears at high temperatures. Taking into account its broad solubility range with $x_{Zn} = 0.37$-$0.56$,[17] a two-sublattice model $(Pd,Zn)_1(Pd,Zn)_1$ is adopted with its Gibbs energy formulated as,

$$G_m = \sum_{i=Pd,Zn}\sum_{j=Pd,Zn} y'_i y''_j \, ^oG_{i:j} + RT\left(\sum_{i=Pd,Zn} y'_i \ln(y'_i) + \sum_{j=Pd,Zn} y''_j \ln(y''_j)\right) \qquad \text{Eq. 8}$$
$$+ y'_{Pd} y'_{Zn}\left(y''_{Pd} L_{Pd,Zn:Pd} + y''_{Zn} L_{Pd,Zn:Zn}\right)$$
$$+ y''_{Pd} y''_{Zn}\left(y'_{Pd} L_{Pd:Pd,Zn} + y'_{Zn} L_{Zn:Pd,Zn}\right)$$

where $y_i^{(s)}$ is the site fraction of component $i$ on sublattice $s$, $^oG_{i:j}$ are the Gibbs energies of the endmembers $(i:j)$, and $L$ are the interaction parameters, which can be expanded using the Redlich-Kister polynomials[33] in the same way as in *Eq. 6* and *Eq. 7*.

FCC_L1$_0$ $\beta_1$ phase is stable at low temperatures (space group $P4/mmm$ with two Wyckoff sties 1a and 1d). A two-sublattice model $(Pd,Zn)_1(Pd,Zn)_1$ is applied for this phase with the Gibbs energy formula similarly to *Eq. 8*.

The $\gamma$ phase has four Wyckoff sites (space group $I\bar{4}3m$) with details presented in Sec.1. A four-sublattice model $(Pd,Zn)_2(Pd,Zn)_3(Pd,Zn)_2(Pd,Zn)_6$ is adopted according to its Wyckoff sites (IT, 8c), (OH, 12e), (OT, 8c), and (CO, 24g), respectively. Gibbs energy of $\gamma$ phase is written as,

$$G_m = \sum_{i=Pd,Zn} \sum_{j=Pd,Zn} \sum_{k=Pd,Zn} \sum_{l=Pd,Zn} y_i' y_j'' y_k''' y_l'''' \, ^oG_{i:j:k:l} \quad \text{Eq. 9}$$

$$+ RT \left( 2 \sum_{i=Pd,Zn} y_i' \ln(y_i') + 3 \sum_{j=Pd,Zn} y_j'' \ln(y_j'') \right.$$

$$\left. + 2 \sum_{k=Pd,Zn} y_k''' \ln(y_k''') + 6 \sum_{l=Pd,Zn} y_l'''' \ln(y_l'''') \right) + \,^{xs}G_m$$

where $y_i^{(s)}$ is the site fraction of component $i$ on sublattice $s$, $s =', '', ''',$ and $''''$ representing the IT, OH, OT, and CO sublattice, respectively, and $^{xs}G_m$ is the excess Gibbs energy. In the present work, we considered only the interaction parameters on the second sublattice (the OH site) due to the lower values of formation enthalpy for the endmembers $(Zn)_2(Pd)_3(Pd)_2(Zn)_6$ and $(Zn)_2(Zn)_3(Pd)_2(Zn)_6$ from DFT-based calculations (Sec.3.1) along with experimental observations[8,21] showing the OH site occupied by both Pd and Zn. Hence, the excess Gibbs energy is as follows

$$^{xs}G_m = y_{Pd}'' y_{Zn}'' y_{Zn}' y_{Pd}''' y_{Zn}'''' L_{Zn:Pd,Zn:Pd:Zn} \quad \text{Eq. 10}$$

In the present work, Pd$_2$Zn, PdZn$_2$, and Pd$_9$Zn$_{91}$ are treated as stoichiometric compounds (phases) with their Gibbs energies expressed as

$$G^{Pd_aZn_b} = e\,^oG_{Pd}^{fcc} + f\,^oG_{Zn}^{hcp} + D + ET \quad \text{Eq. 11}$$

where $^oG_{Pd}^{fcc}$ and $^oG_{Zn}^{hcp}$ are Gibbs energies of FCC-Pd and HCP-Zn, respectively, and $D$ and $E$ are parameters to be evaluated.

Thermodynamic modeling of the Pd-Zn system was carried out by means of the open source software ESPEI,[14] which uses PyCalphad[15] for calculating energies of thermodynamic models. The model parameters are evaluated in two steps in ESPEI. The first step is parameter generation. In this step, the thermochemical data from DFT-based first-principles calculations with all internal degrees of freedom specified,[30,34] such as site fractions in each sublattice, are used to select the number of parameters and evaluate their values. The experimental thermochemical data can also be used in the first step if their internal degrees of freedom are specified, such as stoichiometric compounds or fully random solutions. This is because the minimization of Gibbs energy with respect to the internal variables is not performed in the first step. In the present work, the Gibbs energy functions of stoichiometric compounds and endmembers in the $\beta$, $\beta_1$ and $\gamma$ phases were evaluated from DFT-based first-principles calculations with the results discussed below. In the second step, all model parameters are simultaneously optimized through the Bayesian approach using a Markov Chain Monte Carlo (MCMC) method.[14] The input data for the second step are primarily the experimental phase equilibrium information including two or more co-existing phases and equilibrium thermochemical data.[11–13] The experimental data, including phase boundary data[9,17,18] and thermochemical data,[10,19] were used to refine model parameters. In the present work, each model parameter employed two chains with a standard derivation of 0.1 when initializing in a Gaussian distribution. During the modeling process, the chain values can be tracked and the MCMC steps were performed until the model parameters converged.

The uncertainty quantification of model parameters and calculated thermodynamic properties and phase stability from the models is performed using PDUQ (Phase Diagram Uncertainty Quantification).[16] PDUQ relies on the PyCalphad for predicting thermodynamic properties of interest and ESPEI for Bayesian samples to leverage the distribution of model parameters and estimate uncertainties based on the estimated Gaussian distribution of input data uncertainty.[14,16] The statistical distributions of model parameters are evaluated from the samples during MCMC optimization based on the Metropolis criteria.[14] The values from the last MCMC step were used to estimate the uncertainty. In the present work, 95% uncertainty interval (or Bayesian credible intervals containing 95% of the invariant samples) was applied to quantify the uncertainty.

## 4    Results and discussion

### 4.1    Properties of Pd-Zn compounds by first-principles calculations

Table 3 shows the predicted lattice parameters of $Pd_2Zn$, $PdZn_2$, and γ phase in the present work together with experimental data in the literature.[8,35–37] The lattice parameter c of $Pd_2Zn$ predicted from the present first-principles calculations is 7.83 Å, slightly higher than the experimental 7.65 Å.[35] The lattice parameters of $PdZn_2$ and γ are in good agreement with experimental results with the mean absolute error value around 0.027 Å.

Table 4 shows the equilibrium volume $V_0$, bulk modulus B, and the derivative of bulk modulus B' obtained from the EOS E-V fitting at 0 K in comparison with previous DFT calculations and experimental data.[38] Figure 1 compares phonon DOS's of FCC-Pd, HCP-Zn, and stoichiometric compounds $Pd_2Zn$ and $PdZn_2$. In the low-frequency region (e.g., < 3THz), HCP-Zn has the highest phonon DOS, followed by $PdZn_2$, $Pd_2Zn$, and FCC-Pd. The higher DOS in the low-frequency region results in a lower average phonon frequency.[39] This can be confirmed by the lowest bulk modulus B of HCP-Zn compared with $PdZn_2$, $Pd_2Zn$, and FCC-Pd. The bulk modulus B of HCP-Zn fitted from the present work is 57.5 GPa, which is lower than PdZn$_2$ 118.1 GPa, Pd$_2$Zn 146.0 GPa, and FCC-Pd 167.90 GPa.

Figure 2 and Figure 3 show the comparison of the entropy and enthalpy of FCC-Pd and HCP-Zn from the phonon calculations to the SGTE pure element database.[32] Both show excellent agreement. For results of Pd obtained from phonon calculations and SGTE, the difference of enthalpy is less than 6.5% and that of entropy is less than 5%. The results of Zn show the difference of enthalpy less than 2.1% and that of entropy less than 3.5%.

Table 5 shows the enthalpy of formation $\Delta_f H$ at 0 K predicted from the present DFT-based calculations using the exchange-correlation functionals of GGA and HSE06, along with experimental data.[19,20] For the γ phase, the configuration of Pd$_{10}$Zn$_{42}$ ($x_{Zn}$ = 0.81) was used for DFT-based calculations and compared with experimental data at $x_{Zn}$=0.80. The $\Delta_f H$ value predicted using HSE06 is -33.7 kJ/mol-atom, agreeing reasonably well with -35.1 kJ/mol-atom from experiments using calorimetry.[10] For the $\beta_1$ phase, the configuration of PdZn ($x_{Zn}$ = 0.50) was used. The $\Delta_f H$ value of PdZn predicted using HSE06 is -69.2 kJ/mol-atom, which is in good agreement with measured -73.9±10 kJ/mol-atom reported by Chiang et al.[19] and -66.6 kJ/mol-atom reported by Kou and Chang,[20] but is lower than the value predicted by GGA (-53.7 kJ/mol-atom). For both γ and $\beta_1$ phases, the $\Delta_f H$ values predicted by HSE06 are more accurate than those predicted by PBE-GGA. Considering the high computational cost, HSE06 was only applied for key endmembers close or on the convex hull in the present work such as (Zn)$_2$(Pd)$_3$(Pd)$_2$(Pd)$_6$ and (Zn)$_2$(Zn)$_3$(Pd)$_2$(Zn)$_6$.

### 4.2 Thermodynamic modeling and phase equilibria

Table 6 summarizes the model parameters of the Pd-Zn system from the present work. Figure 4 shows the calculated phase diagram in comparison with experimental data,[9,40,41] showing good agreement, particularly the phase boundaries of the γ phase. The max difference between calculated and experimental Zn compositions of the γ phase $x_{Zn}^{Cal} - x_{Zn}^{Expt.}$ is around 0.016 at 892 K. The solubility range of the γ phase is between $x_{Zn}$ = 0.775-0.846 from 300 K to 1000 K. The congruent melting temperature of the γ phase is 1150 K in good agreement with 1153 K suggested by Massalski.[17]

Table 7 lists the temperatures and compositions of invariant reactions. The experimentally reported invariant reactions, i.e., Liquid → $Pd_9Zn_{91}$ + HCP by Vizdal et al.[9] and other four invariant reactions by Massalski,[17] are well reproduced. The largest discrepancy is seen for the eutectic reaction, Liquid → $Pd_9Zn_{91}$ + HCP, i.e., 681 K calculated from the present work versus 690 K in the literature.[9]

Figure 5 and Figure 6 show the heat capacity, entropy, and enthalpy of stoichiometric compounds $Pd_2Zn$ and $PdZn_2$ from the present model in comparison with results from the phonon calculations. Good agreement is found, especially for enthalpy with a difference less than 4.7% for $Pd_2Zn$ and 2.8% for $PdZn_2$. The entropy of $PdZn_2$ from phonon calculations compared with the present model shows the largest discrepancy with a difference around 6.5 J/mol-atom-K. This is because the model parameters of stoichiometric compounds, which are obtained from first-principles calculated enthalpy and entropy, were adjusted with the experimental data of the peritectoid temperature.

Figure 7 shows the activity values of Zn at 1273 K calculated from the present model in comparison with experimental data by Chiang et al..[19] The activity values of Zn in $\beta_1$ from the present model agree well with the experiments[19] with the mean absolute error of ln $a_{Zn}$ being 0.4. A higher discrepancy occurs in the composition range $x_{Zn}$ = 0.51-0.6. For example, at $x_{Zn}$ = 0.55, ln $a_{Zn}$ calculated from the present model is -3.52 compared with -1.56 measured by Chiang et al..[19] They reported a single $\beta_1$ phase for $x_{Zn}$ = 0.52 and a single $\beta$ phase for $x_{Zn}$ = 0.6. However, the present work predicts that $\beta_1$ is in equilibrium with $\beta$ at $x_{Zn}$ = 0.52, and $\beta$ is in equilibrium with Liquid at $x_{Zn}$=0.6 (see Figure 4). Figure 8 shows the activity values of Zn in FCC, $\beta$, and $\beta_1$ phases, respectively, calculated from the present model with the shaded regions for uncertainty of each phase. For FCC, larger uncertainty occurs when $x_{Zn}$ < 0.2, where the largest uncertainty is around ±15% from the mean value at $x_{Zn}$ = 0.02, and experimental data are located within this uncertainty region. For $\beta_1$, the experimental data are in the uncertainty region when $x_{Zn}$ < 0.5, and the largest error is around $x_{Zn}$ = 0.52 with ln $a_{Zn}$ = -1.66 from experiments[19] compared with -3.35 calculated from the present model. For $\beta$, ln $a_{Zn}$ = -10.33 from experiments[19] at $x_{Zn}$ = 0.3 are closer to the lower limit of the uncertainty region, where ln $a_{Zn}$ at the lower uncertainty is -10.81. The shaded ranges in Figure 8 decrease with increasing $x_{Zn}$, indicating a larger uncertainty of activity occurs in the Pd rich region.

Figure 9 plots the enthalpy of formation of the Pd-Zn phases at 1273 K and 300 K from the present model and available experimental data,[10,19,20] along with the calculated results from the previous

CALPHAD modeling[9] at 300 K and the present first-principles results of the γ phase at 0 K and high temperatures. The enthalpy of formation at 1273 K agrees reasonably well with experimental data.[19,20] The enthalpy of formation of $\beta_1$ at $x_{Zn} = 0.5$ and 1273 K is -70.1 kJ/mol-atom from the present work, compared with -73.9±10 kJ/mol-atom measured by Chiang et al.[19] and -66.6 kJ/mol-atom measured by Kou et al..[20] Figure 9 shows that the enthalpy of formation of γ from the present model has better agreement with experiments than the previous model.[9] At $x_{Zn} = 0.8$, the enthalpy of formation value of γ from the present model is -40.6 kJ/mol-atom at 300 K and from the previous model [9] is -48.5 kJ/mol-atom, compared with -35.1 kJ/mol-atom measured by Amore et al..[10]

## 4.3 Site occupancy in the γ phase and surface construction

Figure 10 shows the calculated site fractions in γ at 773 K and 1023 K from the present model in comparison with XRD results by Edström et al.,[8] Gourdon et al.[21] and Dasgupta et al..[4] Temperature has little influence on the site fraction of the γ phase. With increasing Pd content, Pd is predicted to first occupy the OT sublattice, and then the OH sublattice after the OT sublattice is fully occupied. The site fractions of Pd in the OH sublattice are in good agreement with experimental data. For example, at $x_{Pd} = 0.173$, the calculated site fraction of Pd in the OH sublattice $y_{Pd}^{OH}$ is 0.08, slightly higher than 0.07 measured by Dasgupta et al..[4] At $x_{Pd} = 0.181$, 0.192, and 0.23, the calculated $y_{Pd}^{OH}$ values are 0.118, 0.165, and 0.333, respectively, which agree well with experimental data with mean absolute error of 0.002. The Vizdal et al.[9] model could not predict site fractions in γ due to the 2-sublattice model used.

Force constants can be used to quantitatively understand interactions between atomic pairs.[42] A large and positive force constant suggests strong bonding interaction of an atomic pair, whereas a negative force constant indicates the tendency to separate[43]. To understand site occupancy of Pd in γ, we examined energies and force constants in three $Pd_9Zn_{43}$ configurations to analyze the occupancy of an additional Pd atom compared to the full Pd OT occupation in $Pd_8Zn_{44}$. Three configurations are $(Pd_8)^{OT}(Pd_1Zn_{11})^{OH}(Zn_8)^{IT}(Zn_{24})^{CO}$, $(Pd_8)^{OT}(Zn_{12})^{OH}(Pd_1Zn_7)^{IT}(Zn_{24})^{CO}$, and $(Pd_8)^{OT}(Zn_{12})^{OH}(Zn_8)^{IT}(Pd_1Zn_{23})^{CO}$, where 8 Pd atoms occupy 8 OT sites and the 9th Pd atom occupies one of the OH (conf_OH), IT (conf_IT), or CO (conf_CO) sites, respectively. Force constants can be predicted by DFT-based phonon calculations [44]. Table 2 lists details of phonon calculations for $Pd_9Zn_{43}$ configurations. Figure 11 shows the force constants of atomic pairs Pd-Pd, Pd-Zn, and Zn-Zn in three $Pd_9Zn_{43}$ configurations. The Pd-Zn atomic pairs have the largest force constants 5.373 eV/Å$^2$, compared with 4.875 eV/Å$^2$ of Pd-Pd and 3.569 eV/Å$^2$ of Zn-Zn pairs. It indicates that Pd-Zn pairs has the strongest bonding in the configuration. Table 8 shows the energy and bonding in three $Pd_9Zn_{43}$ configurations, i.e., conf_OH, conf_IT, and conf_CO. conf_OH has the lowest energy $E_{conf\_OH}$ = -105.30 eV/atom and the shortest Pd-Zn bonding distance $d_{conf\_OH}^{Pd-Zn}$ = 2.538 Å in comparison with conf_CO and conf_IT. In contrast, $d_{conf\_OH}^{Pd-Pd}$ = 2.940 Å is larger than that of conf_OH and conf_IT. Table 8 also shows the bonding of the 9th Pd when occupying OH, CO, and IT. In conf_OH, the 9th Pd are bonding with Zn atoms in its first nearest neighbors ($d^{1NN}$ < 3.2 Å, seen in Figure 11), with the nearest Pd atom is 4.713 Å away. In conf_CO and conf_IT, there are Pd-Pd pairs in the first nearest neighbors of the 9th Pd, which make the bonding between Pd and surroundings weaker than that in conf_OH. We conclude that the stability preference for OH occupancy of additional Pd atoms results from stronger Pd-Zn bonding interactions.

Site fractions of γ phase calculated from the CALPHAD modeling are further applied to analyze the surface structures. The possible stable composition of γ phase is evaluated ranging from $Pd_8Zn_{44}$ to $Pd_{12}Zn_{40}$ from the present model. It indicates that site occupancy in OH will be changed with changing Pd composition in γ phase, with OT remained being occupied by Pd, IT and CO occupied by Zn. DFT-based calculations have suggested that the $(1\bar{1}0)$ surface of the γ phase has a lower surface energy than (110) and {111}.[4] Figure 12 shows $(1\bar{1}0)$ surface constructions of γ phase. OT sites separately locate on the surface, forming Pd monomers ($Pd_1$). When increase Pd occupy OH sites, OT-OH-OT assembles on the surface can then become Pd trimers ($Pd_3$). The ability to control the exposure of specific surface ensembles between $Pd_1$ and $Pd_3$ sites is a direct consequence of the site occupancies of the bulk structure and has catalytic consequences. For example, the activity for ethylene hydrogenation and selectivity for acetylene semi-hydrogenation were drastically altered by tuning active ensembles between Pd monomers ($Pd_1$) and Pd trimers ($Pd_3$).[4]

## 5 Conclusions

Thermodynamic modeling of the Pd-Zn system based on the CALPHAD approach has been performed with thermochemical and phase equilibria data from experiments in the literature and DFT-based total energy and phonon calculations in the present work. A 4-sublattice model is used to describe the γ phase in accordance with its four Wyckoff positions providing a better prediction of surface construction and enabling the understanding of active surface ensembles for catalysts. DFT-based first-principles calculations are used to obtain Gibbs energies of endmembers for the γ phase. High throughput CALPHAD modeling tools (i.e., ESPEI and PyCalphad) are employed to evaluate model parameters. Uncertainty of both phase boundaries and activity are quantified in the present work. In the γ phase, Pd atoms first occupy the OT sublattice followed by the OH sublattice, as indicated by DFT-based total energy and phonon calculations supported by the bonding distance and force constants analyses and in agreement with experimental data. Site fractions calculated from the CALPHAD modeling as a function of temperature and composition contributes to analyze surface construction of γ phase, thus implicates to further tailor and understand active assembles on the surface with improved catalytic performance.

**Acknowledgments**

The authors acknowledge the financial support by the Department of Energy (DOE) via Award No. DE-SC0020147. First-principles calculations were performed partially on the Roar supercomputer at the Pennsylvania State University's Institute for Computational and Data Sciences (ICDS), partially on the resources of the National Energy Research Scientific Computing Center (NERSC) supported by the U.S. DOE Office of Science User Facility operated under Contract No. DE-AC02-05CH11231, and partially on the resources of the Extreme Science and Engineering Discovery Environment (XSEDE) supported by NSF with Grant No. ACI-1548562.

**Tables and Table Captions**

Table 1. Crystallographic information for phases in the Pd-Zn system and their sublattice models used in the present CALPHAD modeling.

| Phase name | Strukturbericht | Space group | Pearson symbol | Model |
|---|---|---|---|---|
| Liquid(L) | | | | (Pd,Zn) |
| FCC | A1 | $Fm\bar{3}m$ | cF4 | (Pd,Zn) |
| HCP | A3 | $P6_3/mmc$ | hP2 | (Pd,Zn) |
| BCC_B2($\beta$) | B2 | $Pm\bar{3}m$ | cP2 | $(Pd,Zn)_1(Pd,Zn)_1$ |
| FCC_L1$_0$($\beta_1$) | L1$_0$ | P4/mmm | tP4(tP2) | $(Pd,Zn)_1(Pd,Zn)_1$ |
| Gamma($\gamma$) | D8$_2$ | $I\bar{4}3m$ | cI52 | $(Pd,Zn)_2(Pd,Zn)_3(Pd,Zn)_2(Pd,Zn)_6$ |
| Pd$_2$Zn | | Pnma | oP12(C23) | $(Pd)_2(Zn)_1$ |
| PdZn$_2$ | | Cmmm | oC48(oS48) | $(Pd)_1(Zn)_2$ |
| Pd$_9$Zn$_{91}$ | | | | $(Pd)_{0.09}(Zn)_{0.91}$ |

Table 2. Details of DFT-based first-principles and phonon calculations for each compound or element, including space group, total atom(s) in the cell for the calculations, *k*-points meshes for structure relaxations and the final static calculations (indicated by DFT), supercell sizes for phonon calculations, and *k*-points meshes for phonon calculations.

| Compounds | Space group | Atom(s) in the cell | k-points for DFT | Supercell for phonon | k-points for phonon |
|---|---|---|---|---|---|
| Pd | $Fm\bar{3}m$ | 1 | 22×22×22 | 3×3×3 | 5×5×5 |
| Zn | $P6_3/mmc$ | 2 | 24×24×24 | $\begin{bmatrix} -1 & 2 & 1 \\ -3 & -2 & -1 \\ 1 & -2 & 1 \end{bmatrix}$ | 4×4×4 |
| $Pd_2Zn$ | Pnma | 12 | 12×12×12 | $\begin{bmatrix} -1 & 0 & -1 \\ -1 & 0 & 1 \\ 0 & 2 & 0 \end{bmatrix}$ | 4×4×4 |
| PdZn | P4/mmm | 2 | 19×19×14 | 3×3×3 | 5×5×5 |
| $PdZn_2$ | Cmm2 | 48 | 7×7×4 | 1×1×1 | 4×4×4 |
| $Pd_8Zn_{44}$ | $I\bar{4}3m$ | 52 | 4×4×4 | 1×1×1 | 4×4×4 |
| Endmembers of γ phase | N/A | 26 | 8×8×8 | N/A | N/A |
| $Pd_9Zn_{43}$ [a] | $I\bar{4}3m$ | 52 | 3×3×3 | 1×1×1 | 2×2×2 |

[a] Three $Pd_9Zn_{43}$ configurations used for analysis of site occupancy as discussed in Sec.4.3.

Table 3. Predicted lattice parameters of FCC-Pd, HCP-Zn, $Pd_2Zn$, $PdZn_2$, and γ by first-principles calculations from the relaxed structures at 0 K, together with available experimental (Expt.) data for comparison.

| Phases | a (Å) | b (Å) | c (Å) | Source |
|---|---|---|---|---|
| FCC-Pd | 3.9309 | | | This work |
| | 3.8902 | | | Expt.[45] |
| HCP-Zn | 2.6426 | | 5.0268 | This work |
| | 2.6594 | | 4.9328 | Expt.[46] |
| $Pd_2Zn$ | 5.3975 | 4.1917 | 7.8343 | This work |
| | 5.3500 | 4.1400 | 7.6500 | Expt.[35] |
| $PdZn_2$ | 7.4235 | 7.5773 | 12.3196 | This work |
| | 7.3630 | 7.5250 | 12.3070 | Expt.[41] |
| γ phase | 9.1024 | | | This work |
| | 9.1022 | | | Expt.[8] |
| | 9.0906 | | | Expt.[37] |

Table 4. Equilibrium volume $V_0$, bulk modulus B, and the first derivative of bulk modulus with respect to pressure B', based on the present EOS fittings at 0 K in comparison with the previous DFT studies.

| Phases | $V_0$ (Å³/atom) | B (GPa) | B' | Source |
|---|---|---|---|---|
| Pd | 15.300 | 167.9 | 5.51 | This work |
|  | 15.340 | 163.3 | 5.50 | DFT[38] |
|  | 14.716 | 195.5 |  | Expt.[38] |
| Zn | 15.336 | 57.5 | 5.20 | This work |
|  | 15.491 | 58.6 | 5.01 | DFT[38] |
|  | 15.185 | 73.2 |  | Expt.[38] |
| Pd$_2$Zn | 14.824 | 146.0 | 5.38 | This work |
| PdZn$_2$ | 14.576 | 118.1 | 5.33 | This work |

Table 5. Predicted enthalpy of formation at 0 K, $\Delta_f H$ (kJ/mol-atom), of $\gamma$ and $\beta_1$ using DFT-based calculations with GGA and HSE06 as exchange-correlation functionals, respectively, in comparison with available experimental data.

| Phases | Configurations | $x_{Zn}$ | $\Delta_f H$ | Source |
| --- | --- | --- | --- | --- |
| $\gamma$ phase | $Pd_{10}Zn_{42}$ | 0.81 | -27.9 | DFT/GGA, this work |
|  | $Pd_{10}Zn_{42}$ | 0.81 | -33.7 | DFT/HSE06, this work |
|  | N/A | 0.80 | -35.1 | Calorimetry, at 300 K [10] |
| $\beta_1$ phase | PdZn | 0.50 | -53.7 | DFT/GGA, this work |
|  | PdZn | 0.50 | -69.2 | DFT/HSE06, this work |
|  | N/A | 0.50 | -73.9±10 | Vapor pressure at 1273 K [19] |
|  | N/A | 0.50 | -66.6 | Vapor pressure at 1273 K [20] |

Table 6. Model parameters in the present CALPHAD modeling of the Pd-Zn system with Gibbs energy in J/mol-atom and temperature (T) in Kelvin.

| Phase | Parameters | Value | Ref. |
|---|---|---|---|
| Liquid | $^0L^{Liquid}$ | $-134358 - 7.478\,T$ | |
| | $^1L^{Liquid}$ | $69096$ | |
| | $^2L^{Liquid}$ | $1643 - 5.417\,T$ | |
| FCC_A1 | $^0L^{Fcc}$ | $-207719 + 43.952\,T$ | |
| | $^1L^{Fcc}$ | $4410 + 1\,T$ | |
| | $^2L^{Fcc}$ | $48734 - 6.984\,T$ | |
| HCP | $^0L^{Hcp\_zn}$ | $-134856 - 12.982\,T$ | |
| | $^1L^{Hcp\_zn}$ | $51421$ | |
| BCC_B2($\beta$) | $G^{B2}_{Pd:Pd}$ | $2GHSER_{Pd} + 5000$ | 9 |
| | $G^{B2}_{Pd:Zn}$ | $GHSER_{Pd} + GHSER_{Zn} - 120961 + 16.563\,T$ | |
| | $G^{B2}_{Zn:Pd}$ | $GHSER_{Pd} + GHSER_{Zn} - 120961 + 16.563\,T$ | |
| | $G^{B2}_{Zn:Zn}$ | $2GHSER_{Zn} + 5000$ | 9 |
| | $^0L^{B2}_{Pd:Pd,Zn}={}^0L^{B2}_{Pd,Zn:Pd}$ | $-49594 + 13.968\,T$ | |
| | $^1L^{B2}_{Pd:Pd,Zn}={}^1L^{B2}_{Pd,Zn:Pd}$ | $22460$ | |
| | $^0L^{B2}_{Zn:Pd,Zn}={}^0L^{B2}_{Pd,Zn:Zn}$ | $-94437 + 7.857\,T$ | |
| FCC_L1$_0$($\beta_1$) | $G^{L10}_{Pd:Pd}$ | $2GHSER_{Pd} + 15000$ | |
| | $G^{L10}_{Pd:Zn}$ | $GHSER_{Pd} + GHSER_{Zn} - 134320 + 25.870\,T$ | 9 |
| | $G^{L10}_{Zn:Pd}$ | $GHSER_{Pd} + GHSER_{Zn} - 134320 + 25.870\,T$ | 9 |
| | $G^{L10}_{Zn:Zn}$ | $2GHSER_{Zn} + 15000$ | |

| | | |
|---|---|---|
| | $^0L^{L1_0}_{Pd:Pd,Zn}={}^0L^{L1_0}_{Pd,Zn:Pd}$ | $-35819 + 0.492\,T$ |
| | $^1L^{L1_0}_{Pd:Pd,Zn}={}^1L^{L1_0}_{Pd,Zn:Pd}$ | $28503 + 1.138\,T$ |
| | $^0L^{L1_0}_{Zn:Pd,Zn}={}^0L^{L1_0}_{Pd,Zn:Zn}$ | $-76386 + 3.498\,T$ |
| Gamma(γ) | $G^{Gamma}_{Pd:Pd:Pd:Pd}$ | $13\text{GHSER}_{Pd} + 26802$ |
| | $G^{Gamma}_{Pd:Pd:Pd:Zn}$ | $7\text{GHSER}_{Pd} + 6\text{GHSER}_{Zn} - 83241$ |
| | $G^{Gamma}_{Pd:Pd:Zn:Pd}$ | $11\text{GHSER}_{Pd} + 2\text{GHSER}_{Zn} + 131161$ |
| | $G^{Gamma}_{Pd:Pd:Zn:Zn}$ | $5\text{GHSER}_{Pd} + 8\text{GHSER}_{Zn} - 61982$ |
| | $G^{Gamma}_{Pd:Zn:Pd:Pd}$ | $10\text{GHSER}_{Pd} + 3\text{GHSER}_{Zn} + 186950$ |
| | $G^{Gamma}_{Pd:Zn:Pd:Zn}$ | $4\text{GHSER}_{Pd} + 9\text{GHSER}_{Zn} + 25785$ |
| | $G^{Gamma}_{Pd:Zn:Zn:Pd}$ | $8\text{GHSER}_{Pd} + 5\text{GHSER}_{Zn} + 259218$ |
| | $G^{Gamma}_{Pd:Zn:Zn:Zn}$ | $2\text{GHSER}_{Pd} + 11\text{GHSER}_{Zn} + 190798$ |
| | $G^{Gamma}_{Zn:Pd:Pd:Pd}$ | $11\text{GHSER}_{Pd} + 2\text{GHSER}_{Zn} - 181667$ |
| | $G^{Gamma}_{Zn:Pd:Pd:Zn}$ | $5\text{GHSER}_{Pd} + 8\text{GHSER}_{Zn} - 564558 + 148.359\,T$ |
| | $G^{Gamma}_{Zn:Pd:Zn:Pd}$ | $9\text{GHSER}_{Pd} + 4\text{GHSER}_{Zn} - 64329$ |
| | $G^{Gamma}_{Zn:Pd:Zn:Zn}$ | $3\text{GHSER}_{Pd} + 10\text{GHSER}_{Zn} - 411154$ |
| | $G^{Gamma}_{Zn:Zn:Pd:Pd}$ | $8\text{GHSER}_{Pd} + 5\text{GHSER}_{Zn} - 13907$ |
| | $G^{Gamma}_{Zn:Zn:Pd:Zn}$ | $2\text{GHSER}_{Pd} + 11\text{GHSER}_{Zn} - 440291 + 23.439\,T$ |
| | $G^{Gamma}_{Zn:Zn:Zn:Pd}$ | $6\text{GHSER}_{Pd} + 7\text{GHSER}_{Zn} + 26798$ |
| | $G^{Gamma}_{Zn:Zn:Zn:Zn}$ | $13\text{GHSER}_{Zn} + 88131$ |
|  | $^0L^{Gamma}_{Zn:Pd,Zn:Pd:Zn}$ | $-393082$ |
|  | $^1L^{Gamma}_{Zn:Pd,Zn:Pd:Zn}$ | $-5712$ |
| Pd$_2$Zn | $G^{Pd_2Zn}_{Pd:Zn}$ | $2\text{GHSER}_{Pd} + \text{GHSER}_{Zn} - 142996 + 18.662\,T$ |
| PdZn$_2$ | $G^{PdZn_2}_{Pd:Zn}$ | $\text{GHSER}_{Pd} + 2\text{GHSER}_{Zn} - 160623 + 23.213\,T$ |
| Pd$_9$Zn$_{91}$ | $G^{Pd_{0.09}Zn_{0.91}}_{Pd:Zn}$ | $0.09\text{GHSER}_{Pd} + 0.91\text{GHSER}_{Zn} - 19059 - 0.290\,T$ |

Table 7. Temperatures and compositions of invariant reactions in the Pd-Zn system calculated from the present CALPHAD modeling with available experimental data included.

| Reaction | $x_{Zn}$ | | | Temperature (K) | Ref. |
|---|---|---|---|---|---|
| Liquid → $\beta$ + $\gamma$ | 75.0 | 64.2 | 78.2 | 1123.3 | This work |
|  | 75 | 65 | 77 | 1118 | Expt.[17] |
| $\beta$ → $\beta_1$ + $\gamma$ | 63.2 | 57.3 | 77.5 | 837.6 | This work |
|  | 57 | 55 | 76 | 838 | Expt.[17] |
| $\beta_1$ + $\gamma$ → PdZn$_2$ | 57.2 | 77.5 | 66.7 | 799.6 | This work |
|  | 56 | 76 | 66.7 | 803±10 | Expt.[17] |
| Liquid + $\gamma$ → Pd$_9$Zn$_{91}$ | 97.7 | 84.6 | 91.0 | 707.2 | This work |
|  | 98 | 85 | 92 | 703 | Expt.[18] |
|  |  |  |  | 707 | Expt.[9] |
| Liquid → Pd$_9$Zn$_{91}$ + HCP | 98.2 | 91.0 | 99.0 | 681.1 | This work |
|  |  |  |  | 690 | Expt.[9] |

Table 8. Energies, and distances of bonds (d) for three $Pd_9Zn_{43}$ configurations with the first 8 Pd atoms occupying the OT site, and the 9th Pd atom ($Pd^{9th}$) occupying the OH, CO, or IT site, denoted by conf_OH, conf_CO, and conf_IT, respectively. Configurations are relaxed using DFT calculations.

| Configuration | Energy (eV/atom) | d of Pd-Zn [a] (Å) | d of Pd-Pd [b] (Å) | d of $Pd^{9th}$-Zn [c] (Å) | d of $Pd^{9th}$-Pd [d] (Å) |
|---|---|---|---|---|---|
| conf_OH | -105.30 | 2.538 | 2.940 | 2.563 | 4.713 |
| conf_CO | -104.89 | 2.558 | 2.783 | 2.562 | 2.783 |
| conf_IT | -104.74 | 2.560 | 2.887 | 2.613 | 3.192 |

[a] Distance of the nearest Pd-Zn pair in the configuration (Å).

[b] Distance of the nearest Pd-Zn pair in the configuration (Å).

[c] Distance of the nearest Pd-Zn pair around the 9th Pd atom (Å).

[d] Distance of the nearest Pd-Pd pair around the 9th Pd atom (Å).

**Figures and Figure Captions**

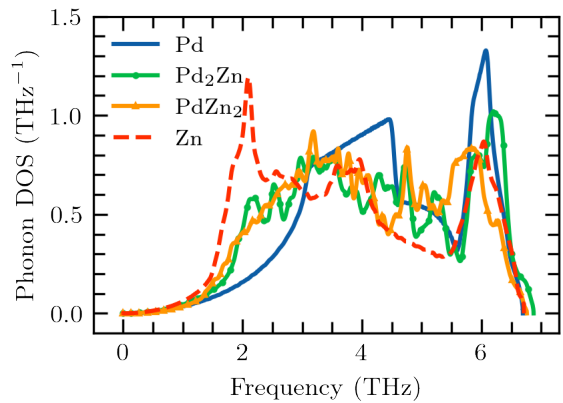

*Figure 1.* Predicted phonon DOS's of Pd, Zn, $Pd_2Zn$, and $PdZn_2$ from the DFT-based phonon calculations.

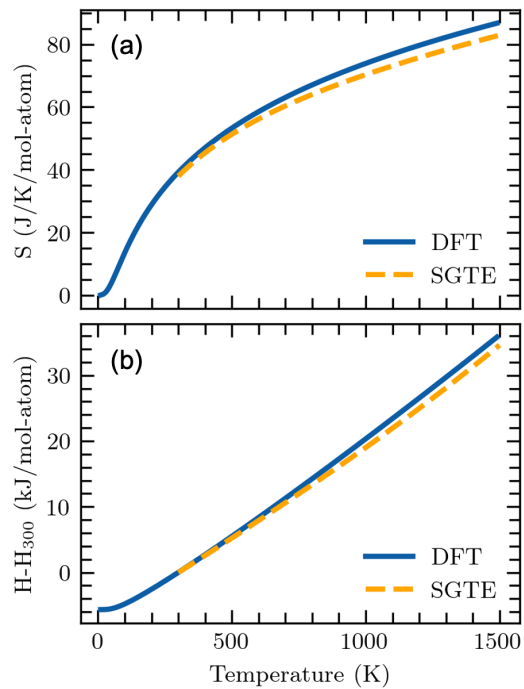

*Figure 2.* Comparison of the (a) entropy and (b) enthalpy of Pd from the DFT-based phonon calculations to the SGTE data.[32]

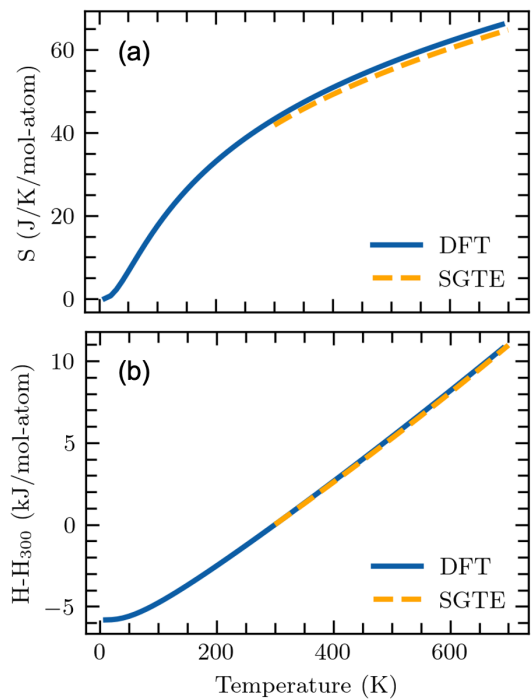

*Figure 3.* Comparison of the entropy and enthalpy of Zn from the phonon calculations to the SGTE data.[32]

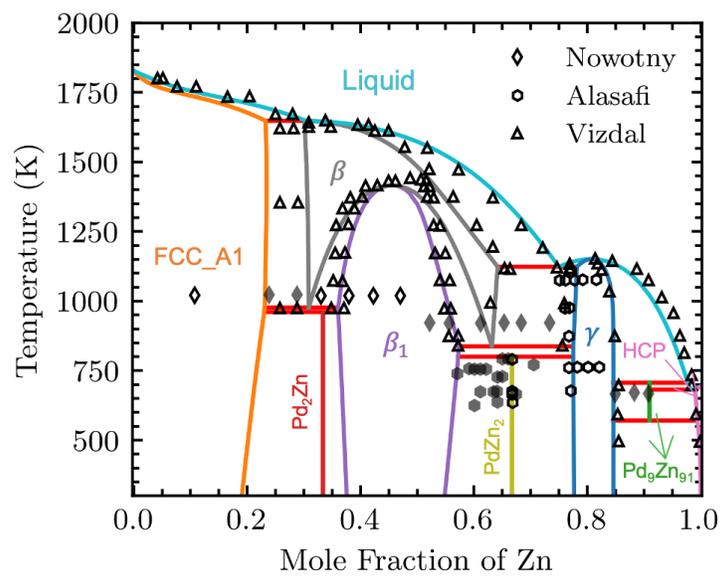

*Figure 4.* Calculated phase diagram from the present CALPHAD modeling in comparison with experimental data from Nowotny et al.[40] (diamonds), Alasafi et al.[41] (hexagons), and experimental data summarized by Vizdal et al.[9] (triangles). Hollow diamonds and hexagons represent single phase region and shadowed diamonds and hexagons represent two phases region reported by Nowotny et al.[40] and Alasafi et al..[41]

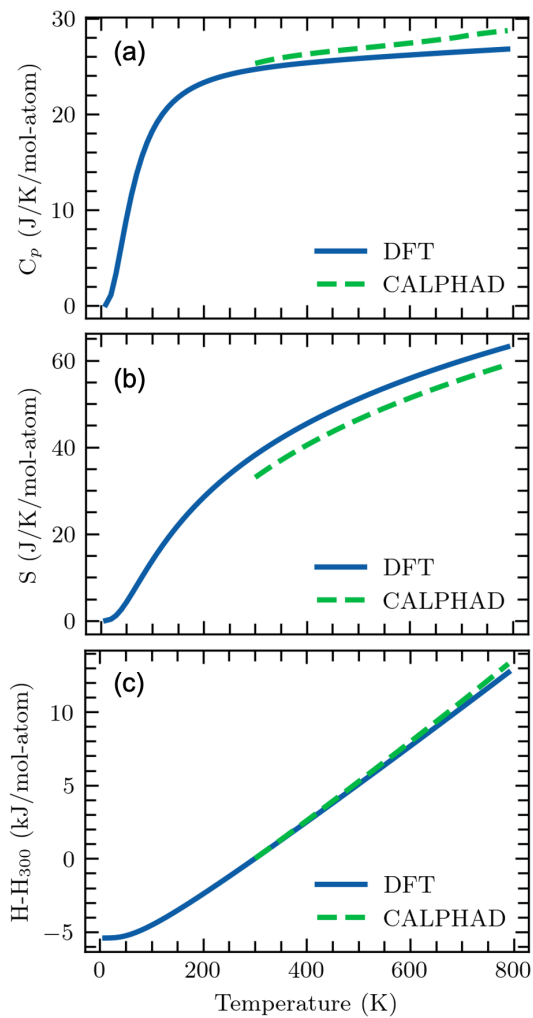

*Figure 5.* Predicted (a) heat capacity, (b) entropy, and (c) enthalpy of $Pd_2Zn$ using the DFT-based phonon calculations, compared with those from the present CALPHAD modeling.

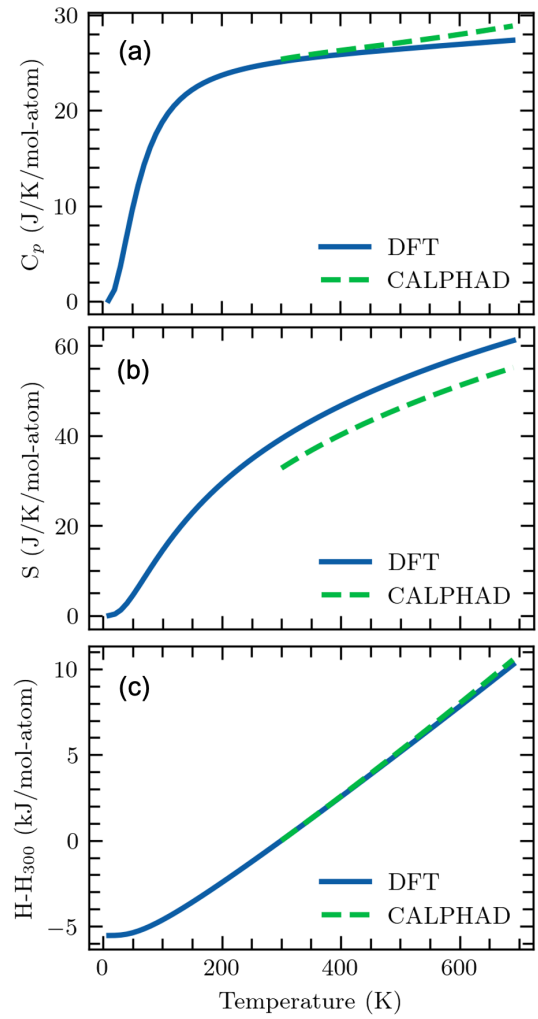

*Figure 6.* Heat (a) capacity, (b) entropy, and (c) enthalpy of PdZn$_2$ using the phonon calculations from first-principles calculations, compared with those from the present CALPHAD modeling.

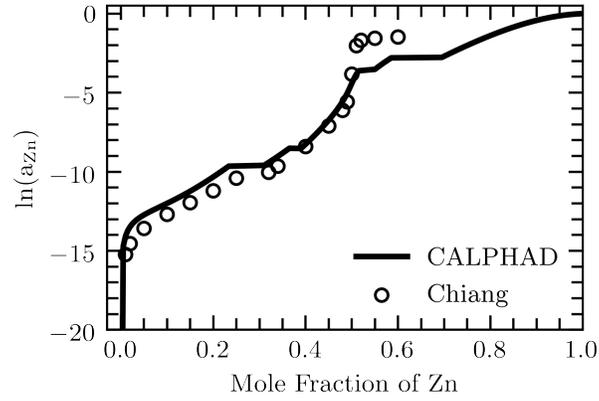

*Figure 7*. Calculated activity of Zn at 1273 K with experimental data measured by Chiang et al.[19] superimposed.

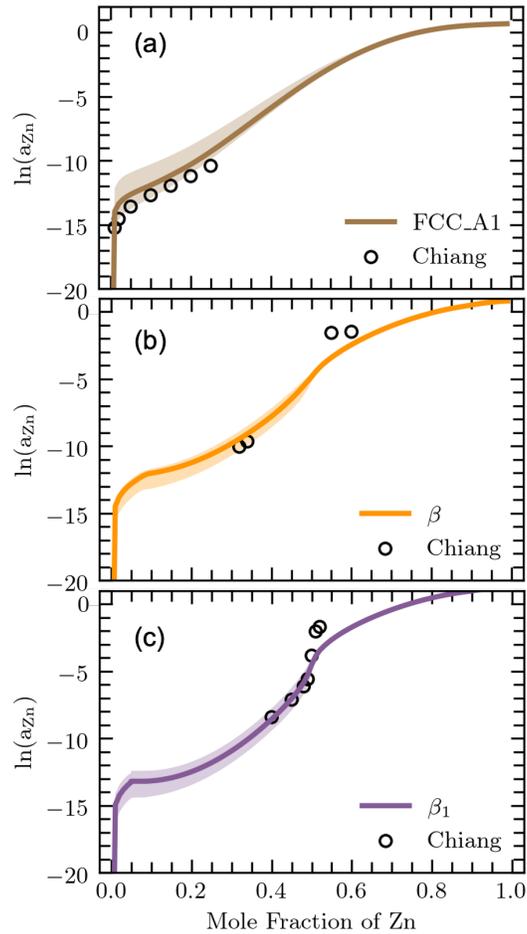

*Figure 8*. Uncertainty quantification of activity of (a) FCC, (c) $\beta$, and (d) $\beta_1$, marked in the shaded regions with corresponding color of each phase.

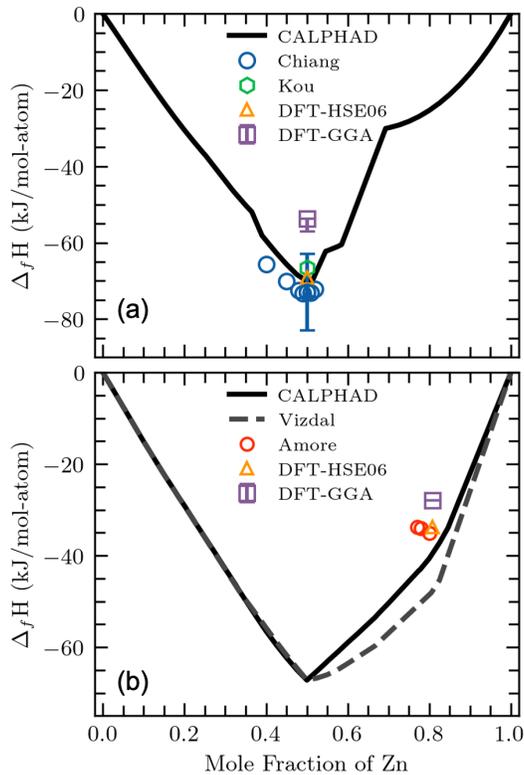

*Figure 9.* Calculated enthalpy of formation at (a) 1273 K and (b) 300 K, along with DFT-based calculations and available experimental data by Chiang et al.,[19] Kou and Chang,[20] and Amore et al..[10] DFT using HSE are calculated at 0 K. DFT using GGA are calculated at 0 K and high temperatures 300 K and 1270 K respectively, showing as purple bars in the figure.

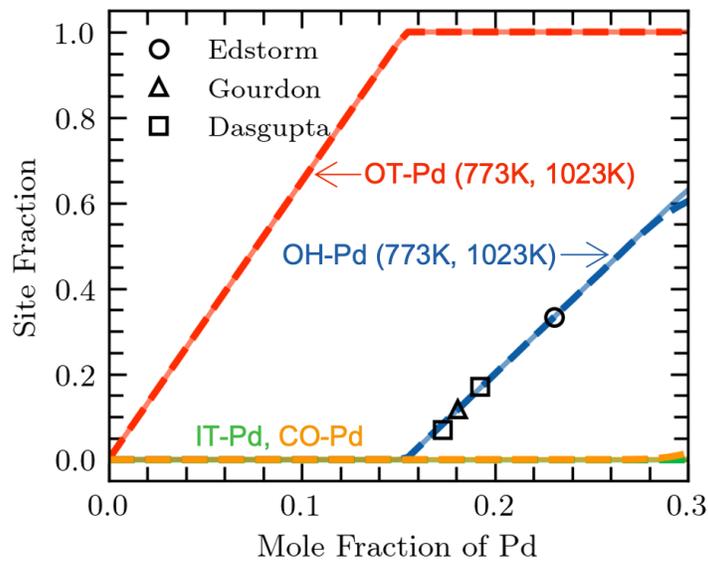

*Figure 10.* Calculated site fractions in γ at 773K (solid lines) and 1023 K (dash lines) with the experimental data by Edström et al.[8] at 923 K, Gourdon et al.[21] at 1023 K, and Dasgupta et al. [4] at 773 K superimposed.

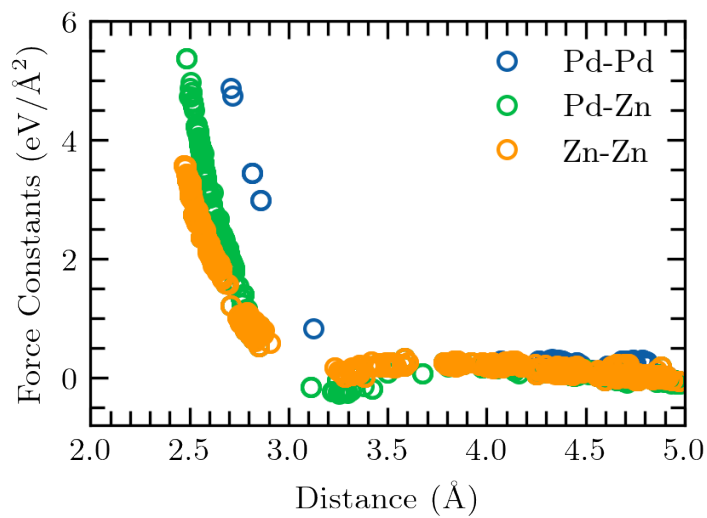

*Figure 11.* Force constants of Pd-Pd, Pd-Zn, and Zn-Zn atom pairs in $Pd_9Zn_{43}$ configurations obtained from phonon calculations.

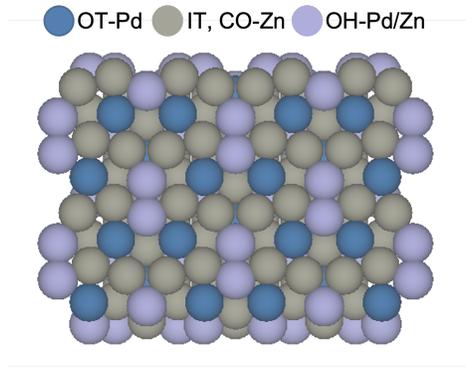

*Figure 12.* (1$\bar{1}$0) surface of γ in 2×2×2 supercell. Blue atoms are OT sites, which are occupied by Pd atoms. Grey atoms are IT and CO sites, which are occupied by Zn atoms. Purple atoms are OH sites, which can be occupied by both Pd and Zn atoms.